\documentclass[pra,twocolumn,amsmath,amssymb,showpacs,superscriptaddress]{revtex4-1}
\usepackage[T1]{fontenc} % for Rzazewski
\usepackage{bm}
\usepackage{graphicx}
\usepackage[bookmarks=false]{hyperref}
\usepackage{color}

\newcommand{\ket}[1]{|#1\rangle}
\newcommand{\braket}[1]{\langle #1 \rangle}
\newcommand{\braketeffg}[1]{\langle #1 \rangle_{\text{g}}^{\text{eff}}}

\def\dd{\mathrm{d}}
\def\ee{\mathrm{e}}
\def\ii{\mathrm{i}}

\def\Re{\mathrm{Re}}

\def\Tr{\mathrm{Tr}}

\def\kB{k_{\text{B}}}
\def\nH{\mathrm{nH}}

\def\fF{\mathrm{fF}}
\def\micron{\mathrm{\mu{}m}}
\def\GHz{\mathrm{GHz}}

\def\ZZ{\mathcal{Z}}
\def\ZZb{\bar{\mathcal{Z}}}
\def\ZZatom{\ZZ_{\text{atom}}}

\def\wc{\omega_{\text{c}}}
\def\wa{\omega_{\text{a}}}
\def\rabi{g}
\def\wab{\tilde{\omega}_{\text{a}}}
\def\rabib{\tilde{g}}
\def\omegab{\tilde{\omega}}
\def\LJb{\tilde{L}_{\text{J}}}
\def\EJb{\tilde{E}_{\text{J}}}

\def\LR{L_{\text{R}}}
\def\CR{C_{\text{R}}}

\def\Zc{Z_{\text{c}}}
\def\Za{Z_{\text{a}}}

\def\Lg{L_{{g}}}

\def\CJ{C_{\text{J}}}
\def\EJ{E_{\text{J}}}
\def\LJ{L_{\text{J}}}

\def\LRz{L_{\text{R}0}}
\def\LRzc{L_{\text{R}0}^{\text{crit}}}

\def\CRz{C_{\text{R}0}}

\def\Zcz{Z_{\text{c0}}}

\def\Zab{\tilde{Z}_{\text{a}}}

\def\oH{\hat{H}}

\def\oHatom{\hat{H}^{\text{atom}}}

\def\oHeff{\hat{H}^{\text{eff}}}

\def\oV{\hat{V}}
\def\oU{\hat{U}}
\def\oUd{\hat{U}^{\dagger}}
\def\oa{\hat{a}}
\def\oad{\hat{a}^{\dagger}}
\def\ob{\hat{b}}
\def\obd{\hat{b}^{\dagger}}
\def\op{\hat{p}}

\def\ophi{\hat{\phi}}
\def\oq{\hat{q}}
\def\opsi{\hat{\psi}}
\def\orho{\hat{\rho}}

\def\ox{\hat{x}}
\def\op{\hat{p}}
\def\oA{\hat{A}}

\def\phiz{\phi_{\text{0}}}
\def\phith{\phi_{\text{eq}}}
\def\psith{\psi_{\text{eq}}}
\def\alphath{\alpha_{\text{eq}}}
\def\fluxq{\Phi_0}
\def\psiext{\Phi_{\text{ext}}}

\begin{document}
\title{Super-radiant phase transition in superconducting circuit
in thermal equilibrium}

\author{Motoaki Bamba}
\altaffiliation{E-mail: bamba@qi.mp.es.osaka-u.ac.jp}
\affiliation{Department of Materials Engineering Science, Osaka University, 1-3 Machikaneyama, Toyonaka, Osaka 560-8531, Japan}
\author{Kunihiro Inomata}
\affiliation{RIKEN Center for Emergent Matter Science (CEMS), 2-1 Hirosawa, Wako, Saitama 351-0198, Japan}
\author{Yasunobu Nakamura}
\affiliation{RIKEN Center for Emergent Matter Science (CEMS), 2-1 Hirosawa, Wako, Saitama 351-0198, Japan}
\affiliation{Research Center for Advanced Science and Technology (RCAST), The University of Tokyo, Meguro-ku, Tokyo 153-8904, Japan}

\date{\today}

\begin{abstract}
We propose a superconducting circuit that shows a super-radiant phase transition (SRPT)
in the thermal equilibrium.
The existence of the SRPT is confirmed analytically in the limit of an infinite number of artificial atoms.
We also perform numerical diagonalization of the Hamiltonian with a finite number of atoms
and observe an asymptotic behavior approaching
the infinite limit as the number of atoms increases.
The SRPT can also be interpreted intuitively in a classical analysis.
\end{abstract}

% PACS, the Physics and Astronomy Classification Scheme.

% 05.30.Rt	Quantum phase transitions
% 42.50.Ct	Quantum description of interaction of light and matter; related experiments
% 85.25.-j	Superconducting devices
% 42.50.Pq	Cavity quantum electrodynamics; micromasers
% 64.70.Tg	Quantum phase transitions (for quantum Hall effects aspects, see 73.43.Nq in electronic structure of surfaces, interfaces, thin films, and low dimensional structures)
% 42.50.-p	Quantum optics

\maketitle

In a variety of studies involving the light-matter interaction,
realization of a super-radiant phase transition (SRPT) still remains
a challenging subject.
It means a spontaneous appearance of coherence amplitude
of transverse electromagnetic fields
due to the light-matter interaction in the thermal equilibrium.
While the laser also shows the spontaneous coherence,
it is generated by population-inverted matters,
i.e., in a non-equilibrium situation.
The SRPT was first proposed theoretically around 1970
\cite{Mallory1969PR,Hepp1973AP,Wang1973PRA},
but afterward its absence in the thermal equilibrium was pointed out
based on the so-called $A^2$ term \cite{Rzazewski1975PRL,Rzazewski1976PRA,Yamanoi1976PLA,Yamanoi1979JPA}
and more generally on the minimal-coupling Hamiltonian
\cite{Bialynicki-Birula1979PRA,Gawedzki1981PRA}.
A SRPT analogue in non-equilibrium situation
was proposed theoretically \cite{Dimer2007PRA}
and was observed experimentally in cold atoms driven by laser light
\cite{Baumann2010N,Baumann2011PRL}.
Realizing a thermal-equilibrium SRPT
and comparing it with the non-equilibrium SRPT (including laser)
are fundamental subjects bridging the statistical physics (thermodynamics),
established in equilibrium situations,
and the electrodynamics (light-matter interaction),
long discussed mostly in non-equilibrium situations.
However, the SRPT has not yet been realized in the thermal equilibrium
since the first proposal \cite{Mallory1969PR,Hepp1973AP,Wang1973PRA}.

\begin{figure}[bp] %%%%%%%%%%%%%%%%%%%%%%%%%%%%%%%%%%%%%%%%%%%%%%%%%%%%%%%%%%%%
\includegraphics[width=\linewidth]{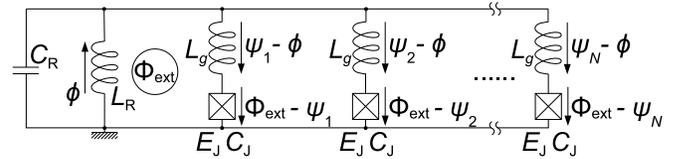}
\caption{Superconducting circuit showing the SRPT in thermal equilibrium
under a static external magnetic flux bias $\psiext=\fluxq/2$.}
\label{fig:1}
\end{figure}

While the atomic systems are basically described
by the minimal-coupling Hamiltonian \cite{Bialynicki-Birula1979PRA,Gawedzki1981PRA},
there are a large number of degrees of freedom in designing
the Hamiltonians of superconducting circuits,
where the existence of the SRPT is still under debate
\cite{Nataf2010NC,Viehmann2011PRL,Ciuti2012PRL,Jaako2016PRA}.
In this Letter, we propose
the superconducting circuit depicted in Fig.~\ref{fig:1}.
We derive the Hamiltonian of this circuit
by the standard quantization procedure \cite{Devoret1997}
as in the recent work which showed the absence of SRPT
in a different circuit structure \cite{Jaako2016PRA}.
We examine its existence in our circuit by using the semi-classical approach
\cite{Wang1973PRA,Hepp1973PRA,Bialynicki-Birula1979PRA,Hemmen1980PLA,Gawedzki1981PRA},
which is known to be justified in the thermodynamic limit (with an infinite number of atoms),
as well as by straightforwardly diagonalizing
the Hamiltonian with a finite number of atoms.

The circuit shown in Fig.~\ref{fig:1} has
a LC resonator with capacitance $\CR$ and inductance $\LR$,
coupled to $N$ parallel branches containing a Josephson junction.
Each junction has Josephson energy $\EJ$ and shunt capacitance $\CJ$
and is connected to the LC resonator through inductance $\Lg$ individually.
This configuration is distinct from the conventional inductive
\cite{Jaako2016PRA,Nataf2010PRL}
and capacitive couplings \cite{Jaako2016PRA},
where the existence of the SRPT was proposed \cite{Nataf2010NC,Ciuti2012PRL}
but was denied afterward \cite{Viehmann2011PRL,Jaako2016PRA}.
However, the no-go result was shown only for such specific configurations \cite{Viehmann2011PRL,Jaako2016PRA}
and is not applied to ours.
% The SRPT in our circuit is enabled by the competition of the inductive energies
% in the junctions and the resonators as discussed below.
We first explain why the SRPT occurs in our circuit
by analyzing the form of the Hamiltonian.

We apply a static external flux bias $\psiext=\fluxq/2$
in the loop between the resonator and the junctions,
where $\fluxq=h/(2e)$ is the flux quantum.
Alternatively, we can remove the external field
and replace the Josephson junctions with $\pi$ junctions
which have an inverted energy-phase relation \cite{Ryazanov2001PRL}.
We define the ground and the branch fluxes $\phi$ and $\{\psi_j\}$
($j=1,\ldots,N$) as in Fig.~\ref{fig:1}.
According to the flux-based procedure \cite{Devoret1997},
the Hamiltonian is derived straightforwardly and quantized as
\begin{equation} \label{eq:oH_EJcos} % !!!!!!!!!!!!!!!!!!!!!!!!!!!!!!!!!!!!!!!
\oH = \frac{\oq^2}{2\CR} + \frac{\ophi^2}{2\LR}
  + \sum_{j=1}^N \left[
      \frac{\orho_j{}^2}{2\CJ} + \frac{(\opsi_j-\ophi)^2}{2\Lg}
    + \EJ\cos\frac{2\pi\opsi_j}{\fluxq}
    \right].
\end{equation}
Here, $\oq$ and $\{\orho_j\}$ are the conjugate momenta of
$\ophi$ and $\{\opsi_j\}$, respectively, satisfying
$[\ophi,\oq]=\ii\hbar$
and $[\opsi_j,\orho_{j'}]=\ii\hbar\delta_{j,j'}$.

\begin{figure}[tbp] %%%%%%%%%%%%%%%%%%%%%%%%%%%%%%%%%%%%%%%%%%%%%%%%%%%%%%%%%%%%
\includegraphics[width=\linewidth]{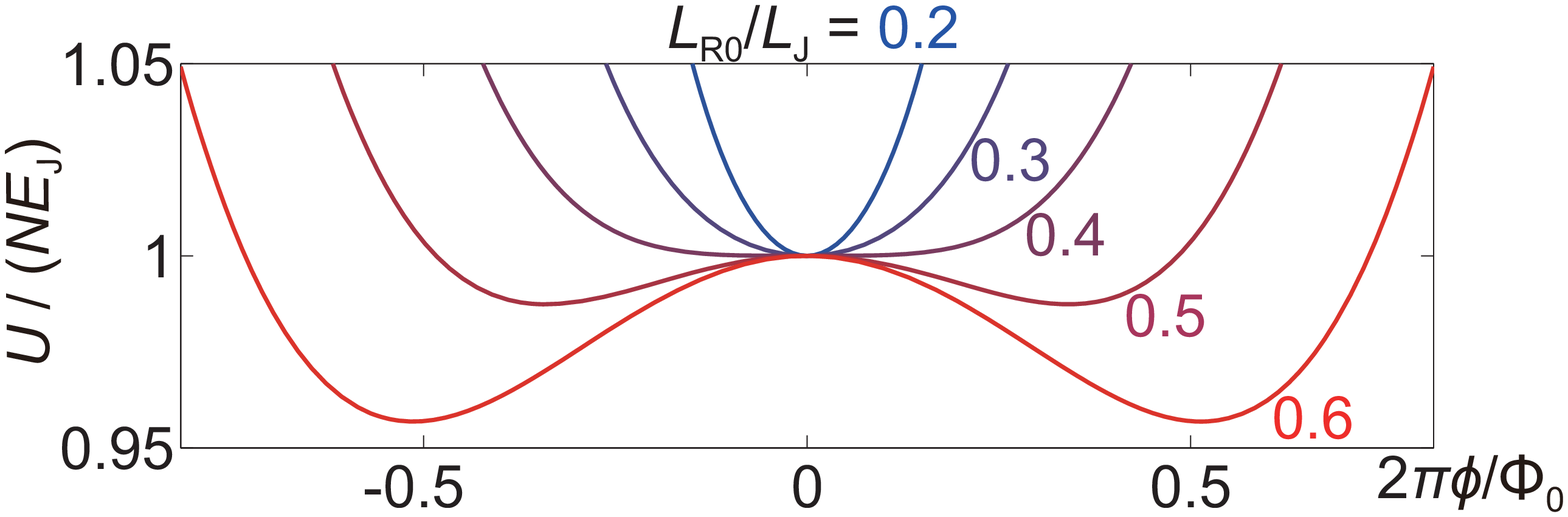}
\caption{Normalized inductive energy [Eq.~\eqref{eq:U(phi,psi)}]
versus $2\pi\phi/\fluxq$ under the condition of $\psi_j=(1+\Lg/\LRz)\phi$,
which is obtained by $\partial{}U/\partial\phi=0$.
For $N\LR=\LRz>\LJ-\Lg$, 
the inductive energy shows two minima at $\phi\neq0$.
This transition corresponds to the SRPT in the sense of the quantum phase transition.
Parameter: $\Lg=0.6\LJ$.
}
\label{fig:2}
\end{figure}
Let us first understand intuitively the SRPT in our circuit by a classical analysis.
The inductive energy in Eq.~\eqref{eq:oH_EJcos} is extracted as
\begin{equation} \label{eq:U(phi,psi)} % !!!!!!!!!!!!!!!!!!!!!!!!!!!!!!!!!!!!!
U(\phi, \{\psi_j\})
= \frac{\phi^2}{2\LR}
  + \sum_{j=1}^N \left[
      \frac{(\psi_j-\phi)^2}{2\Lg}
    + \EJ\cos\frac{2\pi\psi_j}{\fluxq}
    \right].
\end{equation}
The energy minima correspond to the ground state in the classical physics.
While the parabolic terms
$\phi^2/(2\LR)$ and $(\psi_j-\phi)^2/(2\Lg)$
are minimized at $\phi=\psi_j=0$,
the anharmonic term $\EJ\cos(2\pi\psi_j/\fluxq)$
is minimized at $\psi_j=\pm\fluxq/2$.
This is owing to the external flux bias $\psiext=\fluxq/2$;
the sign of the last term
in Eqs.~\eqref{eq:oH_EJcos} and \eqref{eq:U(phi,psi)} is positive,
because the phase difference across the Josephson junction is
given by $\pi-2\pi(\psi_j/\Phi_0)$
due to the flux quantization in each loop
consisting of $\LR$, $\Lg$, and the junction.
This competition between the parabolic and anharmonic inductive energies
is the trick for realizing the SRPT.

For simplifying the following discussion,
we define an inductance $\LJ\equiv[\fluxq/(2\pi)]^2/\EJ$ by the Josephson energy $\EJ$.
The inductive energy $U$ in Eq.~\eqref{eq:U(phi,psi)}
is minimized for $\psi_j=[1+\Lg/(N\LR)]\phi$,
which is obtained from $\partial{}U/\partial\phi=0$.
Since the SRPT is basically discussed in the thermodynamic limit $N\to\infty$,
we scale the inductance of the LC resonator
by the number $N$ of junctions as
\begin{equation} \label{eq:LR_N} % !!!!!!!!!!!!!!!!!!!!!!!!!!!!!!!!!!!!!!!!!
\LR = \LRz/N,
\end{equation}
where $\LRz$ is $N$-independent inductance.
Then, the inductive energy $U/N$ per junction becomes independent of $N$.
In Fig.~\ref{fig:2},
we plot $U/(N\EJ)$ as a function of $2\pi\phi/\fluxq$
under the condition of $\psi_j=(1+\Lg/\LRz)\phi$.
The five curves in Fig.~\ref{fig:2} are the results
for different $\LRz$ under a fixed $\Lg$ of $0.6\LJ$.
The inductive energy $U$ is minimized at $\phi=\psi_j=0$
for $\LRz<\LJ-\Lg=0.4\LJ$,
since the parabolic terms dominate.
In contrast, the anharmonic term dominates
when $\LRz$ satisfies
\begin{equation} \label{eq:cond_SRPT} % !!!!!!!!!!!!!!!!!!!!!!!!!!!!!!!!!!!!!!
N\LR = \LRz > \LJ - \Lg.
\end{equation}
Then, $U$ is minimized at the two points
with non-zero fluxes $\phi=\pm\phiz$.
In the quantum theory, the real ground state is a superposition
of the two minimum points (conceptually speaking,
$\ket{\text{g}}=\frac{\ket{\phiz}+\ket{-\phiz}}{\sqrt{2}}$)
and the expectation values of the fluxes are zero
$\braket{\text{g}|\ophi|\text{g}}=\braket{\text{g}|\opsi_j|\text{g}}=0$
for finite $N$.
However, the thermodynamic limit $N\to\infty$ justifies the classical approach,
since the height of the potential barrier in the whole system is proportional to $N$.
Then, in this limit, we find a spontaneous appearance of coherence
(symmetry breaking), i.e., non-zero $\phi=\pm\phi_0$
and $\psi_j=\pm(1+\Lg/\LRz)\phi_0$ appear in the circuit.
This transition from $\phi=0$ to $\pm\phiz$ by changing $\LRz$
corresponds to the SRPT
in the sense of the quantum phase transition
\cite{Emary2003PRL,Emary2003PRE}
as discussed below in a quantum analysis.

Let us compare Eq.~\eqref{eq:oH_EJcos} with the minimal-coupling Hamiltonian
(under the long-wavelength approximation as discussed in Ref.~\cite{Bialynicki-Birula1979PRA})
\begin{equation} \label{eq:oHmin} % !!!!!!!!!!!!!!!!!!!!!!!!!!!!!!!!!!!!!!!!!!
\oH_{\text{min}} = \oH_{\text{em}} + \sum_{j}^{N} \frac{(\op_j-e\oA)^2}{2m} + \oV(\{\ox_j\}).
\end{equation}
Here, $\oH_{\text{em}}$ represents the energy of the transverse electromagnetic fields
described by the vector potential $\oA$ and its conjugate momentum.
The second and the last terms are, respectively,
the kinetic and the Coulomb interaction energies
of particles with mass $m$, charge $e$, and momentum $\{\op_j\}$
at position $\{\ox_j\}$.
The kinetic energy $(\op_j-e\oA)^2/(2m)$
corresponds to the inductive one $(\opsi_j-\ophi)^2/(2\Lg)$ at $\Lg$
in Eq.~\eqref{eq:oH_EJcos}.
In this way, $\opsi_j$ and $\ophi$ correspond to $\op_j$ and $\oA$,
respectively,
and then $\orho_j$ corresponds to $\ox_j$.
% The essential difference between Eqs.~\eqref{eq:oH_EJcos} and \eqref{eq:oHmin}
% is the description of the anharmonicity.
The no-go theorem of the SRPT in the minimal-coupling Hamiltonian
relies on the fact that the mixing term $(\op_j-e\oA)^2/(2m)$
and the anharmonic term $\oV(\{\ox_j\})$
are, respectively, described by $\op_j$ and $\ox_j$
\cite{Bialynicki-Birula1979PRA,Gawedzki1981PRA}.
In contrast,
in our Hamiltonian, Eq.~\eqref{eq:oH_EJcos},
both the mixing $(\opsi_j-\ophi)^2/(2\Lg)$
and the anharmonicity $\EJ\cos(2\pi\opsi_j/\fluxq)$
are described by $\opsi_j$.
This is the essence for avoiding the no-go theorem
\cite{Bialynicki-Birula1979PRA,Gawedzki1981PRA}
and also for the transition discussed in the above classical analysis.
In the Supplemental Material \cite{suppl},
we also explain how we avoid the no-go results based on the $A^2$ term
\cite{Rzazewski1975PRL,Rzazewski1976PRA,Yamanoi1976PLA,Yamanoi1979JPA}
and on the $P^2$ one
\cite{Emeljanov1976PLA,Yamanoi1978P2}.
The latter corresponds to the direct qubit-qubit interaction
discussed recently in Ref.~\cite{Jaako2016PRA}.

By decomposing $\sum_{j=1}^N(\opsi_j-\ophi)^2/(2\Lg)$ in Eq.~\eqref{eq:oH_EJcos},
we obtain $N\ophi^2/(2\Lg)$.
This corresponds to the $A^2$ term \cite{Rzazewski1975PRL,Rzazewski1976PRA}
(since $\ophi$ corresponds to $\oA$)
and renormalizes the frequency of the LC resonator
as $\wc=\sqrt{({N}/{\Lg}+{1}/{\LR})/\CR}$.
Here, in order to make $\wc$ independent of $N$,
in addition to the scaling of $\LR$ in Eq.~\eqref{eq:LR_N},
we also scale the capacitance $\CR$ as
\begin{equation} \label{eq:CR_N} % !!!!!!!!!!!!!!!!!!!!!!!!!!!!!!!!!!!!!!!!!
\CR = N \CRz,
\end{equation}
where $\CRz$ is $N$-independent capacitance.
Introducing an annihilation operator
$\oa\equiv\ophi/\sqrt{2\hbar\Zc}+\ii\oq\sqrt{\Zc/(2\hbar)}$,
where the impedance $\Zc$ is scaled as $\Zc=\Zcz/N$
for $\Zcz=\sqrt{({1}/{\Lg}+{1}/{\LRz})^{-1}/\CRz}$,
the Hamiltonian in Eq.~\eqref{eq:oH_EJcos} is rewritten as
\begin{equation} \label{eq:oH=ada+Hatom} % !!!!!!!!!!!!!!!!!!!!!!!!!!!!!!!!!!!
\oH = \hbar\wc\left(\oad\oa+\frac{1}{2}\right) - \frac{\ophi}{\Lg}\sum_{j=1}^N\opsi_j
+ \sum_{j=1}^N \oHatom_j.
\end{equation}
Here, the Hamiltonian involving the $j$-th junction is
\begin{equation} \label{eq:oHatom} % !!!!!!!!!!!!!!!!!!!!!!!!!!!!!!!!!!!!!!!!!
\oHatom_j = \frac{\orho_j{}^2}{2\CJ} + \frac{\opsi_j{}^2}{2\Lg}
    + \EJ\cos\frac{2\pi\opsi_j}{\fluxq}.
\end{equation}
Although we cannot obtain $\oHatom_j$
by simply extracting a part of elements from the circuit in Fig.~\ref{fig:1},
this anharmonic oscillator described by $\opsi_j$ and $\orho_j$
is formally considered to be our ``atom''.
The first term in Eq.~\eqref{eq:oH=ada+Hatom} is the Hamiltonian
of our ``photons'', which is described by $\ophi$ and $\oq$ or $\oa$,
renormalized by the $A^2$ term.
The second term in Eq.~\eqref{eq:oH=ada+Hatom} is our ``photon-atom interaction''.
In the following, we discuss the SRPT in terms of these photons and atoms
in relationship to the past discussions on the Dicke model
\cite{Mallory1969PR,Hepp1973AP,Wang1973PRA,Ciuti2012PRL,Nataf2010NC,Emary2003PRL,Emary2003PRE}.

In contrast to the two-level atoms considered in the Dicke model,
our atoms have weakly-nonlinear bosonic transitions
(we explain in detail the parameters used in the following calculations
in the Supplemental Material \cite{suppl}).
In Fig.~\ref{fig:3}(b),
the solid curve represents
the atomic wavefunction at each atomic level,
which are calculated from $\oHatom_j$ in Eq.~\eqref{eq:oHatom}.
The dashed curve represents the inductive energy as a function of $2\pi\psi_j/\fluxq$.
In this Letter,
we basically consider the case $\LJ>\Lg=0.6\LJ$.
Then, since the anharmonic energy $\EJ\cos(2\pi\psi_j/\fluxq)$
is smaller than the parabolic one $\psi_j{}^2/(2\Lg)$,
as seen in Fig.~\ref{fig:3}(b),
the inductive energy in each atom is minimized at $\psi_j=0$,
and the nonlinearity is only about $3\%$.

Here, we tentatively neglect the anharmonicity
as $\EJ\cos(2\pi\opsi_j/\fluxq)\simeq\EJ-\opsi_j{}^2/(2\LJ)$
and assume that the transitions in each atom are described approximately
by bosonic annihilation operator
$\ob_j=\opsi_j/\sqrt{2\hbar\Za}+\ii\orho_j\sqrt{\Za/(2\hbar)}$ as
\begin{align} \label{eq:oHzero} % !!!!!!!!!!!!!!!!!!!!!!!!!!!!!!!!!!!!!!!!!
\oH
& \simeq \hbar\wc\left(\oad\oa+\frac{1}{2}\right)
  + \sum_{j=1}^N\hbar\wa\left(\obd_j\ob_j+\frac{1}{2}\right)
\nonumber \\ & \quad
  - \frac{\hbar\rabi}{\sqrt{N}}(\oa+\oad)\sum_{j=1}^N (\ob_j+\obd_j) + N\EJ.
\end{align}
Here, $\wa=\sqrt{({1}/{\Lg}-{1}/{\LJ})/\CJ}$ %=2\pi\times31.7\;\GHz$
is the atomic transition frequency,
and the interaction strength $\rabi$ is expressed as
$\rabi={\sqrt{N\Zc\Za}}/{2\Lg}={\sqrt{\Zcz\Za}}/{2\Lg}$,
where we define another impedance
$\Za=\sqrt{({1}/{\Lg}-{1}/{\LJ})^{-1}/\CJ}$. %=217\;\Omega$.}

Two transition frequencies $\omega_{\pm}$ of the bosonized Hamiltonian in Eq.~\eqref{eq:oHzero}
are obtained easily by the Bogoliubov transformation \cite{hopfield58,Ciuti2005PRB} as
\begin{equation} \label{eq:w_polariton} % !!!!!!!!!!!!!!!!!!!!!!!!!!!!!!!!!!!!
\omega_{\pm}{}^2 = \frac{\wc{}^2+\wa{}^2 \pm \sqrt{(\wc{}^2-\wa{}^2)^2+16\rabi^2\wc\wa}}{2}.
\end{equation}
Note that $\wc$, $\wa$, $\rabi$, and $\omega_{\pm}$ are not scaled with $N$
by the scaling of $\LR$ and $\CR$ in Eqs.~\eqref{eq:LR_N}
and \eqref{eq:CR_N}, respectively.
The expression in Eq.~\eqref{eq:w_polariton} is exactly the same
as the one derived in the Holstein-Primakoff approach
for the Dicke model \cite{Emary2003PRL,Emary2003PRE}.
When $\omega_-$ becomes imaginary,
the normal ground state (showing $\braket{\text{g}|\oa|\text{g}}=\braket{\text{g}|\ophi|\text{g}}=0$) becomes unstable.
However, it is an artifact due to the neglect of the anharmonicity
in Eq.~\eqref{eq:oHzero}.
As found in the classical analysis in Fig.~\ref{fig:2},
the real ground state appears at an inductive energy minimum
with $\phi=\pm\phi_0$ (super-radiant ground state
with a photonic amplitude of $\braket{\text{g}|\oa|\text{g}}\simeq\pm\phi_0/\sqrt{2\hbar\Zc}$)
in the presence of the anharmonicity
\cite{Yamanoi1976PLA,Yamanoi1979JPA,Emary2003PRL,Emary2003PRE,Bamba2014SPT}.
Equation \eqref{eq:w_polariton} suggests that
the super-radiant ground state appears for $4\rabi^2>\wc\wa$, 
which gives exactly the same condition as Eq.~\eqref{eq:cond_SRPT}
obtained by the classical approach.
While these conditions are certainly satisfied in our circuit,
we will obtain a more rigorous condition of the SRPT
in the following semi-classical analysis.

The above classical and bosonic quantum analyses
imply the SRPT in the sense of the quantum phase transition,
i.e., in the limit of $T\to0$. On the other hand,
the thermodynamic properties at a finite temperature $T$
is analyzed by the partition function
$\ZZ(T)=\Tr[\ee^{-\oH/\kB T}]$.
Since we have only one photonic mode in our system, 
in the thermodynamic limit $N\to\infty$,
the trace over the photonic states
can be replaced by the integral over the coherent state
\cite{Wang1973PRA,Hepp1973PRA,Bialynicki-Birula1979PRA,Hemmen1980PLA,Gawedzki1981PRA}
(see also the Supplemental Material \cite{suppl}) as
\begin{equation} \label{eq:Z(T)=int_alpha} % !!!!!!!!!!!!!!!!!!!!!!!!!!!!!!!!!!
\ZZ(T)
\simeq \int\frac{\dd^2\alpha}{\pi}\
    \ee^{-{\hbar\wc(|\alpha|^2+1/2)}/{\kB T}} \ZZatom(\alpha,T)^N.
\end{equation}
Here, $\alpha\in\mathbb{C}$ is the amplitude of the coherent state $\ket{\alpha}$
giving $\oa\ket{\alpha}=\alpha\ket{\alpha}$.
The atomic partition function
$\ZZatom(\alpha,T)=\Tr_j[\ee^{-\oHeff_j(\alpha)/\kB{}T}]$
is defined with an effective Hamiltonian for a given $\alpha$ as
\begin{equation} \label{eq:Heff} % !!!!!!!!!!!!!!!!!!!!!!!!!!!!!!!!!!!!!!!!!!!
\oHeff_j(\alpha)
= - \frac{1}{\Lg}\phi(\alpha)\opsi_j
  + \oHatom_j,
\end{equation}
where the flux amplitude is expressed as
% \begin{equation}
$\phi(\alpha)=\sqrt{2\hbar\Zc}\Re[\alpha]=\sqrt{2\hbar\Zcz}\Re[\alpha]/\sqrt{N}$.
% \end{equation}
The partition function in Eq.~\eqref{eq:Z(T)=int_alpha} is rewritten as
$\ZZ(T)\simeq\int({\dd^2\alpha}/{\pi}) \ee^{-S(\alpha,T)/\kB{}T}$,
where the action is expressed as
$S(\alpha,T)=\hbar\wc(|\alpha|^2+1/2)-N\kB{}T\ln\ZZatom(\alpha,T)$.
Since the photon-atom interaction is mediated by $\phi(\alpha)$ in Eq.~\eqref{eq:Heff},
the minimum action is obtained for $\alpha\in\mathbb{R}$.
Then, the photonic amplitude $\alphath$
in the thermal equilibrium at $T$
is determined by $\dd{}S(\alpha,T)/\dd\alpha=0$,
which is rewritten as
\begin{equation} \label{eq:wc_alpha_g_beta} % !!!!!!!!!!!!!!!!!!!!!!!!!!!!!!!!
\left(\frac{1}{\LRz}+\frac{1}{\Lg}\right)\phith
- \frac{1}{\Lg}\frac{\Tr\left[ \opsi_j\ee^{-\oHeff_j(\alphath)/\kB T} \right]}{\ZZatom(\alphath,T)}
= 0,
\end{equation}
where $\phith=\phi(\alphath)$ is the flux amplitude.

\begin{figure}[tbp] %%%%%%%%%%%%%%%%%%%%%%%%%%%%%%%%%%%%%%%%%%%%%%%%%%%%%%%%%%
\includegraphics[width=\linewidth]{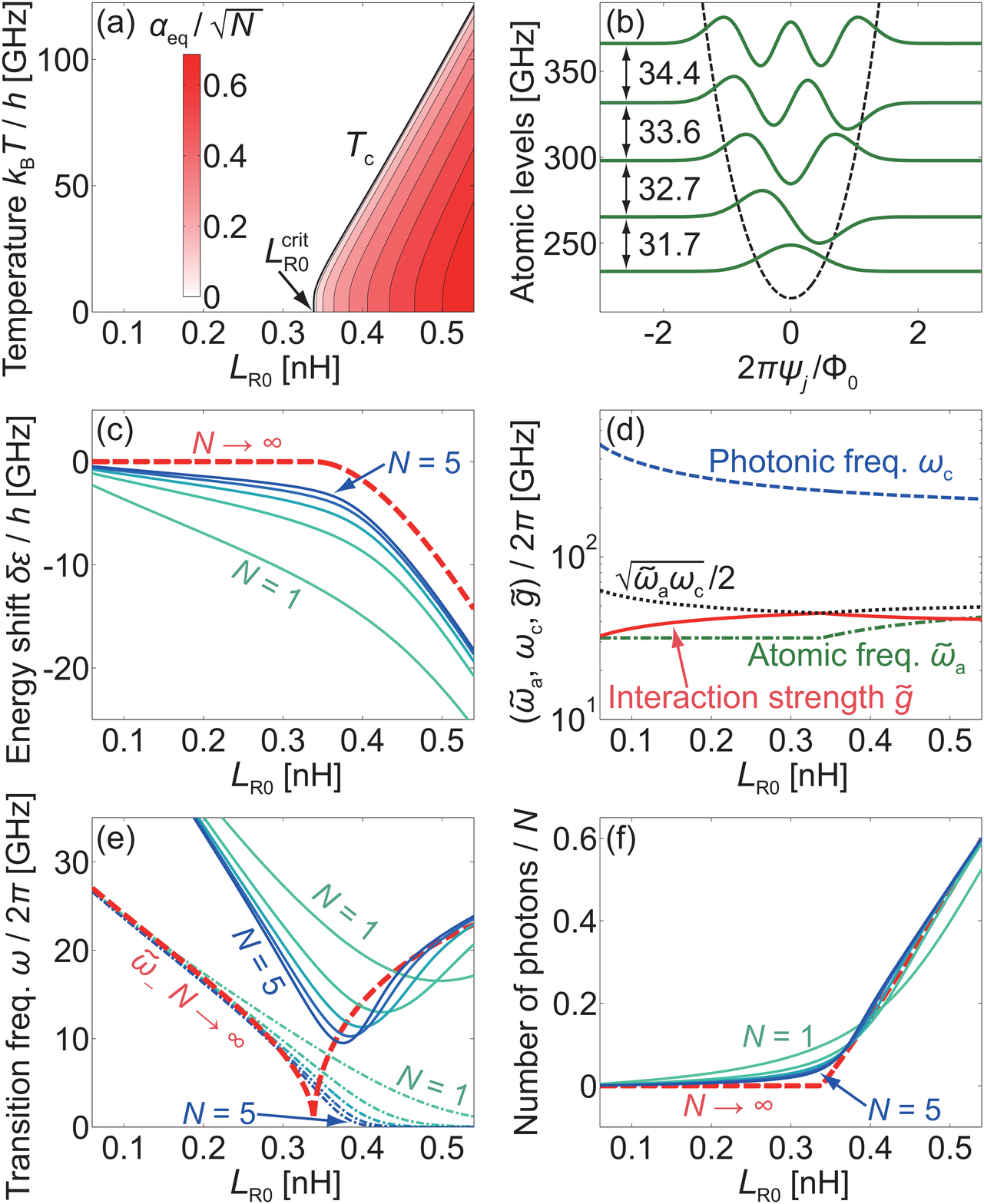}
\caption{
(a) Normalized photonic amplitude $\alphath/\sqrt{N}$
in the thermodynamic limit (infinite number of atoms $N\to\infty$)
versus $\LRz=N\LR$ and temperature $\kB T/h$ in frequency unit.
The SRPT occurs at $\LRzc$ at $T = 0$.
(b) Atomic levels, potential, and wavefunctions.
(c) Zero-point energy shift,
(e) transition frequency,
and (f) number of photons per atom
versus $\LRz$.
Dashed curves are obtained in the thermodynamic limit,
and solid curves are calculated for $N=1,\ldots,5$.
In (e), the dash-dotted curves represent the transition frequencies
for exciting odd number of bosons,
while the solid curves are those for exciting even number of bosons.
(d) Photonic frequency $\wc$, renormalized atomic frequency $\wab$,
and renormalized light-matter interaction strength $\rabib$
versus $\LRz$.
The SRPT occurs when $\rabib$ reaches
the critical interaction strength $\sqrt{\wab\wc}/2$
plotted by the dotted line.
Parameters: $\LJ=0.75\;\nH$, $\Lg=0.6\LJ=0.45\;\nH$,
$\CJ=24\;\fF$, and $\CRz=2\;\fF=\CR/N$.
}
\label{fig:3}
\end{figure}
In Fig.~\ref{fig:3}(a),
the normalized photonic amplitude $\alphath/\sqrt{N}$
calculated by Eq.~\eqref{eq:wc_alpha_g_beta} is color-plotted
as a function of the inductance $\LRz=N\LR$ and temperature $\kB T/h$ in frequency unit.
What is important is the ratio of the inductances.
The Josephson inductance is assumed to be
$\LJ=[\fluxq/(2\pi)]^2/\EJ=0.75\;\nH$,
and the connecting inductance $\Lg=0.6\LJ=0.45\;\nH$.
At $T=0$,
the non-zero photonic amplitude $\alphath$ appears
for $\LRz>\LRzc\simeq0.34\;\nH$,
which roughly agrees with the classical result $\LJ-\Lg=0.30\;\nH$
in Eq.~\eqref{eq:cond_SRPT}. The deviation is
due to the quantum treatment of atoms in Eq.~\eqref{eq:wc_alpha_g_beta}.
The photonic flux $\phith$ roughly agrees
with $\phi_0$ giving the minimum of the inductive energy $U$ in Eq.~\eqref{eq:U(phi,psi)}
in the classical analysis.
The non-zero $\alphath$ and $\phith$ appear even at finite temperatures,
and the critical temperature $T_{\text{c}}$ increases
with increasing $\LRz$.
This is the main evidence of the thermal-equilibrium SRPT in our circuit,
which is obtained in the thermodynamic limit $N\to\infty$.

The transition frequencies,
which are experimentally observable, are obtained
by quantizing the fluctuation around the equilibrium value $\phith$ 
determined by Eq.~\eqref{eq:wc_alpha_g_beta}.
A similar analysis has been performed
in Refs.~\cite{Emary2003PRL,Emary2003PRE}
for the Dicke Hamiltonian.
The atomic flux at $T=0$ is determined
as the expectation value $\psith=\braketeffg{\opsi_j}$
in the ground state of the effective Hamiltonian $\oHeff_j(\alphath)$.
The original Hamiltonian in Eq.~\eqref{eq:oH_EJcos} is expanded
with respect to the fluctuations $\delta\ophi=\ophi-\phith$ and
$\delta\opsi_j=\opsi_j-\psith$
up to $O(\delta\opsi_j{}^2)$ \cite{suppl} as
\begin{align} \label{eq:oH_delta_phi_psi} % !!!!!!!!!!!!!!!!!!!!!!!!!!!!!!
\oH
& \simeq \frac{\oq^2}{2\CR} + \frac{\delta\ophi^2}{2\LR}
+ \sum_{j=1}^N \left\{
    \frac{\orho^2}{2\CJ} + \frac{(\delta\ophi-\delta\opsi_j)^2}{2\Lg}
  - \frac{\delta\opsi_j{}^2}{2\LJb}
  \right\}
\nonumber \\ & \quad
 + N (\EJ + \delta\varepsilon).
\end{align}
Here,
$\EJb=\EJ\braketeffg{\cos(2\pi\opsi/\fluxq)}$
and $\LJb = [\fluxq/(2\pi)]^2/\EJb$
are modified by the quantum treatment of atoms even before the SRPT.
The zero-point energy shift per atom is
\begin{equation} \label{eq:delta_E0} % !!!!!!!!!!!!!!!!!!!!!!!!!!!!!!!!!!!!!!!
\delta\varepsilon = \frac{\phith{}^2}{2\LRz} + \frac{(\phith-\psith)^2}{2\Lg}
    + \EJb-\EJ.
\end{equation}
This shift $\delta\varepsilon/h$ is plotted as the dashed curve in Fig.~\ref{fig:3}(c).
After the appearance of non-zero amplitudes $\phith$ and $\psith$,
the zero-point energy is decreased by the light-matter interaction.

The transition frequencies $\omegab_{\pm}$ are calculated also
from Eq.~\eqref{eq:w_polariton}
but with replacing $\wa$ and $\rabi$
by $\wab=[({1}/{\Lg}-{1}/{\LJb})/\CJ]^{1/2}$
and $\rabib={[\Zcz\Zab}]^{1/2}/{(2\Lg)}$, respectively,
where $\Zab=[({1}/{\Lg}-{1}/{\LJb})^{-1}/\CJ]^{1/2}$.
$\wab$, $\wc$, and $\rabib$ are plotted in Fig.~\ref{fig:3}(d).
When $\rabib$ reaches the critical value $\sqrt{\wab\wc}/2$ plotted by the dotted line,
the SRPT occurs.
Compared with the simple condition $4\rabi^2>\wc\wa$
(giving the critical inductance $\LJ-\Lg=0.30\;\nH$)
obtained from Eq.~\eqref{eq:oHzero} under the bosonic approximation,
the deviation is due to the consideration of the anharmonicity
in Eq.~\eqref{eq:oH_delta_phi_psi}.
The lower transition frequency $\omegab_-$
is plotted by the dashed line in Fig.~\ref{fig:3}(e).
It never becomes imaginary value but shows a cusp at $\LRzc$.
In contrast to the Dicke model \cite{Emary2003PRL,Emary2003PRE},
$\omegab_-$ does not become zero even at $\LRzc$ in our system.
It is due to the presence of multiple atomic levels with anharmonicity
(our atom cannot be equivalent with the two-level limit
since $\LJ>\Lg$ is required).
On the other hand, in the limit of negligible anharmonicity ($\LJ\gg\Lg$),
the SRPT condition in Eq.~\eqref{eq:cond_SRPT} is justified,
and $\omegab_-$ drops to zero at $\LRzc$.

Here, we note that, in the limit of $\Lg\to0$,
we definitively get $\psi_j=\phi$ as seen in Eq.~\eqref{eq:oH_EJcos}
or in the circuit of Fig.~\ref{fig:1},
while the transition in Fig.~\ref{fig:2} itself occurs in the classical approach.
We also get $\wa,\rabi\to\infty$,
while $4\rabi^2>\wc\wa$ is reduced to $\LRz>\LJ$,
and the transition still remains.
However, in order to distinguish the photons and atoms
and to discuss the SRPT, we need a finite $\Lg$.
Especially, as far as we checked numerically,
$\LRz$, $\LJ$, and $\Lg$ are desired to be in the same order
to observe a sharp drop of the transition frequency
for finite $N$ as we will see in the following.

In addition to the above semi-classical approach
justified in the thermodynamic limit $N\to\infty$,
we also diagonalize numerically the original Hamiltonian
in Eq.~\eqref{eq:oH=ada+Hatom} for $N=1,2,\ldots,5$
in order to predict the tendency to be observed in experiments
with a finite number of junctions.
For expressing the Hamiltonian with a sparse matrix
and reducing the computational cost,
we expand $\cos(2\pi\opsi_j/\fluxq)$
and consider the terms up to $O(\opsi_j{}^4)$
(see the details of the numerical diagonalization
in the Supplemental Material \cite{suppl}).

Since the total Hamiltonian in Eq.~\eqref{eq:oH_EJcos}
or \eqref{eq:oH=ada+Hatom} has the parity symmetry,
the expectation value of the photonic amplitude in the ground state is basically zero for finite $N$.
The non-zero $\alphath$ is obtained only in the thermodynamic limit
$N\to\infty$.
In the numerical diagonalization of the Hamiltonian,
we first consider the subsystem with even numbers of bosons
as in Ref.~\cite{Emary2003PRE},
because it is not mixed with the other subsystem with odd numbers of bosons.
For the obtained ground state $\ket{\text{g}}$ with an energy $E_{\text{g}}$,
the expectation number of photons $\braket{\text{g}|\oad\oa|\text{g}}/N$ per atom
is plotted in Fig.~\ref{fig:3}(f) for $N=1,\ldots,5$.
The dashed curve shows $\alphath{}^2/N$ in the thermodynamic limit.
Figure \ref{fig:3}(c) shows the zero-point energy shift
$\delta\varepsilon=(E_{\text{g}}-\hbar\wc/2)/N-\varepsilon_{\text{a}0}$ per atom,
where $\varepsilon_{\text{a}0}$ is the atomic zero-point energy seen in Fig.~\ref{fig:3}(b).
The transition frequency from the ground state
to the first excited state in the even-number subsystem
is plotted as solid curves in Fig.~\ref{fig:3}(e).
It sharply drops around $\LRzc$ even with $N=5$ atoms
and asymptotically reproduces the thermodynamic limit (dashed curve)
for $\LRz>\LRzc$.
On the other hand, the dash-dotted curves represent the transition frequency
from the ground state to the lowest state in the odd-number subsystem.
They asymptotically approach the thermodynamic limit for
$\LRz<\LRzc$ and vanish for $\LRz>\LRzc$ as seen also in Refs.~\cite{Emary2003PRE,Nataf2010PRL}.
These characteristic features imply the existence of the SRPT,
i.e., the parity symmetry breaking \cite{Emary2003PRL,Emary2003PRE}.

We conclude that the superconducting circuit in Fig.~\ref{fig:1}
shows the SRPT in the thermal equilibrium.
It was confirmed by the semi-classical approach in the thermodynamic limit.
It was also checked by calculating the number of photons,
transition frequency, and zero-point energy shift
in the numerical diagonalization of the Hamiltonian with a finite number of atoms.
Experimentally, the transition frequency could be observed
by measuring the excitation spectra,
which would reveal a drastic behavior around the critical point,
as seen in Fig.~\ref{fig:3}(e),
by changing $\LRz$, by decreasing the temperature,
or by increasing the number of atoms.

\begin{acknowledgments}
We thank P.-M.~Billangeon for fruitful discussions.
M.~B.~also thanks F.~Yoshihara for critical comments.
This work was funded by ImPACT Program of Council for Science, Technology and
Innovation (Cabinet Office, Government of Japan)
and by JSPS KAKENHI (Grants No.~26287087, 26220601, and 15K17731).
\end{acknowledgments}

\section*{Supplemental Material}
\appendix
In the main text,
we explain why the SRPT occurs in our circuit
by analyzing the form of the Hamiltonian,
i.e., we show that the transition of the inductive energy minima corresponds to the SRPT
and that the anharmonic term described by $\opsi_j$ is essential in our SRPT.
While we consider that they are the most intuitive and general explanations,
we also explain how we avoid the no-go results based on the $A^2$ term
in Appendix \ref{app:A2} and on the $P^2$ term in Appendix \ref{app:P2},
where we also show that the direct qubit-qubit interaction is obtained
by a unitary transformation of our Hamiltonian
and corresponds to the $P^2$ term.
The justification of the semi-classical analysis is shown in
Appendix \ref{app:semi}.
The derivation of the effective Hamiltonian around the equilibrium
is performed in Appendix \ref{app:fluc}.
In Appendix \ref{app:diag},
we show the details of the numerical diagonalization
of the Hamiltonian for a finite number of junctions.

\section{Why the $A^2$ term does not prevent our SRPT} \label{app:A2}
While the $A^2$ term prevents the SRPT in the atomic systems
\cite{Rzazewski1975PRL,Rzazewski1976PRA,Yamanoi1976PLA,Yamanoi1979JPA}
and also in the superconducting circuits with the conventional capacitive coupling
\cite{Viehmann2011PRL},
it does not prevent it in our system.
Here we explain how we avoid the no-go result based on the $A^2$ term.
In our Hamiltonian, the inductive energy at $\Lg$ is expanded as
\begin{equation}
\sum_{j=1}^N\frac{(\opsi_j-\ophi)^2}{2\Lg}
= \frac{N\ophi^2}{2\Lg}
- \sum_{j=1}^N\frac{\ophi\opsi_j}{\Lg}
+ \sum_{j=1}^N\frac{\opsi_j{}^2}{2\Lg}.
\end{equation}
The first term is the $A^2$ term,
and the second term represents the photon-atom interaction.
The third term is a part of the atomic Hamiltonian as
\begin{equation} \label{eq:oHatom2} % !!!!!!!!!!!!!!!!!!!!!!!!!!!!!!!!!!!!!!!!!
\oHatom_j = \frac{\orho_j{}^2}{2\CJ} + \frac{\opsi_j{}^2}{2\Lg}
    + \EJ\cos\frac{2\pi\opsi_j}{\fluxq}.
\end{equation}
As discussed in the main text,
$\opsi_j{}^2/(2\Lg)$ corresponds to the kinetic energy $\op_j{}^2/(2m)$
of charged particles.
The essential difference from the minimal-coupling Hamiltonian is
that our atomic Hamiltonian in Eq.~\eqref{eq:oHatom2}
has the anharmonic term $\EJ\cos(2\pi\opsi_j/\fluxq)$
described by $\opsi_j$ as we also mention in the main text.
As the result, our particle mass $\Lg$
is effectively increased to $(1/\Lg-1/\LJ)^{-1}$
owing to the anharmonic term.
Then, the atomic transition frequency 
\begin{equation}
\wa=\sqrt{({1}/{\Lg}-{1}/{\LJ})/\CJ}
\end{equation}
is lowered
and the interaction strength 
\begin{equation}
\rabi=\sqrt{\Zcz\sqrt{({1}/{\Lg}-{1}/{\LJ})^{-1}/\CJ}}/{(2\Lg)}
\end{equation}
is increased, 
compared respectively with $\sqrt{1/(\Lg\CJ)}$ and
$\sqrt{\Zcz\sqrt{\Lg/\CJ}}/{(2\Lg)}$
obtained in the absence of the anharmonic term.
In contrast, the $A^2$ term $N\ophi^2/2\Lg$ and $\wc$
are not modified by the anharmonic term.
As a result, the interaction strength $\rabi$
can exceed the critical value $\sqrt{\wc\wa}/2$ for the SRPT in our system.
This is the reason why the $A^2$ term does not prevent our SRPT.

\section{Direct junction-junction interaction and the SRPT} \label{app:P2}
The absence of SRPT in the superconducting circuits with the conventional capacitive coupling
was discussed in terms of the direct qubit-qubit interaction in Ref.~\cite{Jaako2016PRA},
which corresponds to the $P^2$ term \cite{Emeljanov1976PLA,Yamanoi1978P2}
in the atomic systems as we will explain later.
In the same manner as the unitary transformation
between the Hamiltonians with the $A^2$ and the $P^2$ terms
\cite{Keeling2007JPCM,Bamba2014SPT},
we can transform our Hamiltonian to another expression
with a direct junction-junction interaction as in Ref.~\cite{Jaako2016PRA}.

By introducing a unitary operator
\begin{equation}
\oU = \exp\left( \frac{1}{\ii\hbar} \ophi \sum_{j=1}^N \orho_j\right),
\end{equation}
the flux $\opsi_j$ in each junction and the charge $\oq$ in the LC resonator
are transformed as
\begin{subequations}
\begin{align}
\oUd\opsi_j\oU & = \opsi_j + \ophi, \\
\oUd\oq\oU & = \oq - \sum_{j=1}^N \orho_j.
\end{align}
\end{subequations}
Then, our Hamiltonian is transformed as
\begin{align} \label{eq:oUHU} % !!!!!!!!!!!!!!!!!!!!!!!!!!!!!!!!!!!!!!!!!!!!
\oUd\oH\oU & = \frac{1}{2\CR}\left(\oq - \sum_{j=1}^N \orho_j\right)^2
   + \frac{\ophi^2}{2\LR}
\nonumber \\ & \quad
  + \sum_{j=1}^N \left[
      \frac{\orho_j{}^2}{2\CJ} + \frac{\opsi_j}{2\Lg}
    + \EJ\cos\frac{2\pi(\opsi_j+\ophi)}{\fluxq}
    \right].
\end{align}
Expanding the first term, we get
$\sum_{j,j'}\orho_j\orho_{j'}/(2\CR)$,
which corresponds to the direct qubit-qubit interaction discussed in Ref.~\cite{Jaako2016PRA}.
Further, since $\orho_j$ corresponds to the position $\ox_j$ of the charged particle,
this term also corresponds to the $P^2$ term,
the square of the electric polarization
$\hat{P}(x) = \sum_{j}e\ox_j\delta(x-\ox_j)$ in atomic systems
as discussed in Refs.~\cite{Emeljanov1976PLA,Yamanoi1978P2}.
In the same manner as the $A^2$ term,
this $P^2$ term prevents the SRPT in atomic systems
\cite{Emeljanov1976PLA,Yamanoi1978P2}
and also in superconducting circuits with the conventional capacitive coupling
\cite{Jaako2016PRA}.

However, since our Hamiltonian has the anharmonic term
$\EJ\cos(2\pi\opsi_j/\fluxq)$ described by $\opsi_j$,
we get again the $A^2$ term (and also $A^4$, $A^6$, \ldots terms)
from the last term in Eq.~\eqref{eq:oUHU}
after the unitary transformation.
As discussed in Refs.~\cite{Knight1978PRA,Keeling2007JPCM},
the SRPT occurs in some Hamiltonians with both the $A^2$ and $P^2$ terms.
While it is hard to compare Eq.~\eqref{eq:oUHU}
with the Hamiltonians in Refs.~\cite{Knight1978PRA,Keeling2007JPCM}
due to the presence of the $A^4$, $A^6$, \ldots terms,
the SRPT occurs in our system as discussed in the main text
and also in Appendix \ref{app:A2}.
In this way, the anharmonic term described by $\psi_j$ is essential for our SRPT.

\section{Justification of the semi-classical approach} \label{app:semi}
In the semi-classical approach,
we replace the trace over the photonic states
by the integral over the coherent state in Eq.~(11) of the main text.
The justification of this replacement has been discussed in 
Refs.~\cite{Wang1973PRA,Hepp1973PRA,Hemmen1980PLA,Gawedzki1981PRA}.
In the early study by Wang and Hioe \cite{Wang1973PRA},
they noted that this replacement is justified
on the following two assumptions
\cite{Wang1973PRA}:
\begin{enumerate}
\item The limits as $N\rightarrow\infty$ of the field operator $\oa/\sqrt{N}$
and $\oad/\sqrt{N}$ exist.
\item The order of the double limit in the exponential series
$\lim_{N\rightarrow\infty}\lim_{R\rightarrow\infty}\sum_{r=1}^R (-\beta \oH)^r/r!$
can be interchanged.
\end{enumerate}
The validities of these assumptions have been discussed
for some atomic models including the case
with a finite number of the electromagnetic modes \cite{Hepp1973PRA}.
While the first assumption seems to be satisfied
since $\alphath/\sqrt{N}$ ($\propto \phith$) shows a finite value after the SRPT
in the thermodynamic limit $N\to\infty$,
it is hard to show whether our system satisfies the second assumption.
Instead, we justify our semi-classical approach
according to the discussion in Refs.~\cite{Hepp1973PRA}.
The exact partition function $\ZZ(T)=\Tr[\ee^{-\oH/\kB T}]$
and the approximated one $\ZZb(T)$ in Eq.~(11) of the main text
show the following relation \cite{Hepp1973PRA}:
\begin{equation}
\ZZb(T) \leq \ZZ(T) \leq \exp\left(\frac{1}{\kB T}\sum_{k=1}^M\hbar\omega_k\right)\ZZb(T).
\end{equation}
Here, we generally consider $M$ photonic modes with frequencies $\{\omega_k\}$,
while our system has only $M=1$ mode with $\wc$.
The free energy per atom is
\begin{multline}
-\frac{1}{N} \sum_{k=1}^M\hbar\omega_k
-\frac{\kB T}{N}\ln\ZZb(T)\\
\leq -\frac{\kB T}{N}\ln\ZZ(T)
\leq  -\frac{\kB T}{N}\ln\ZZb(T).
\end{multline}
Then, if the number $M$ of photonic modes is finite
($\lim_{N\to\infty}\sum_{k=1}^M\hbar\omega_k/N=0$),
and the free energy $-({\kB T}/{N})\ln\ZZb(T)$ per atom is also a finite value,
$\ZZ(T)$ is well approximated by $\ZZb(T)$
in the thermodynamic limit.
In our case, we numerically checked that
the free energy per atom $-(\kB T/N)\ln\ZZb(T)$ converges to a finite value
with increasing the number of atomic levels considered in the calculation.
Thus, the replacement performed in Eq.~(11) is justified.

\section{Deriving the effective Hamiltonian around equilibrium} \label{app:fluc}
For deriving the effective Hamiltonian
around the equilibrium [Eq.~(14) in the main text],
we expand the anharmonic term in Eq.~(1) of the main text
up to $O[(\delta\opsi_j/\fluxq)^2]$ as
\begin{align}
\cos(2\pi\opsi_j/\fluxq)
& \simeq \braketeffg{\cos(2\pi\opsi_j/\fluxq)}
  [ 1 - (2\pi\delta\opsi_j/\fluxq)^2/2 ]
\nonumber \\ & \quad
  - \braketeffg{\sin(2\pi\opsi_j/\fluxq]}(2\pi\delta\opsi_j/\fluxq).
\end{align}
Further, we use $(1/\LRz+1/\Lg)\phith - \psith/\Lg = 0$
obtained from Eq.~(13) at zero temperature
and $(\psith-\phith)/\Lg - (2\pi/\fluxq)\EJ\braketeffg{\sin(2\pi\opsi_j/\fluxq)} = 0$
obtained from $[\orho_j, \oHeff_j(\alpha)] = 0$.
% From these relations, in the main text, Eq.~(14) is obtained from Eq.(1).

\section{Numerical diagonalization of Hamiltonian for a finite number of junctions} \label{app:diag}
For expressing the Hamiltonian with a sparse matrix
and reducing the computational cost,
the atomic Hamiltonian in Eq.~(8) of the main text is approximated as
\begin{align} \label{eq:oHatom_b_psi4} % !!!!!!!!!!!!!!!!!!!!!!!!!!!!!!!!!!
\oHatom_j
& \simeq \hbar\wa\left(\obd_j\ob_j+\frac{1}{2}\right)
  + \frac{\EJ}{4!} \left(\frac{2\pi\opsi_j}{\fluxq}\right)^4 + \EJ.
\end{align}
Each atom is represented in the Fock basis up to 24 bosons,
which is large enough since the expectation number of bosons
is at most 3.25 under the parameters in Fig.~3 of the main text.
The influence of this approximation (truncating
higher Taylor series) was checked numerically.
We found at most 1.5\% difference between the lowest eight transition frequencies
obtained for the original atomic Hamiltonian in Eq.~(8) of the main text
and for the approximated one in Eq.~\eqref{eq:oHatom_b_psi4}.

The parameters used in this study are chosen
basically for the computational convenience,
and we can choose other values
without changing our results qualitatively.
We find that 
$\CJ=24\;\fF$, $\LJ=0.75\;\nH$ ($\EJ/h=218\;\GHz$), and $\Lg=0.6\LJ$
suppress sufficiently the expectation number of bosons in each atom
and also the influence of the truncation performed
in Eq.~\eqref{eq:oHatom_b_psi4}.
These values are realized by a Josephson junction
with an area of around $S=0.4\;\micron^2$
and a relatively large inductance structure for $\Lg$.
They were searched numerically under the restrictions that
each atom is made by a Josephson junction
and three inductances $\LJ$, $\Lg$, and $\LRzc$ are in the same order,
by which the drop of the transition frequency is sharpened even for $N=5$
as seen in Fig.~3(e) of the main text.
In order to suppress the expectation number of photons,
we need to consider $\wc \gg \wa$.
For this reason, we set $\CRz = 2\;\fF$,
while our results are not changed qualitatively
even if $\CRz$ has a different magnitude.

In the numerical diagonalization of the Hamiltonian
for the finite number of atoms
under the truncation in Eq.~\eqref{eq:oHatom_b_psi4},
the maximum number of bosons is restricted to 48 in the whole system
(and maximally 24 bosons in each mode)
in both calculations for the even-number and odd-number subsystems.
We numerically checked that the results converge sufficiently
by this Fock basis.

We also calculated the transition frequencies
for $N = 1$ and 2 without the truncation
(but the maximum number of bosons is only 16 in each mode
due to the computational restriction).
Note that the transition frequency from the ground state
to the first excited state in the even-number subsystem
[solid curves in Fig.~3(e) of the main text] clearly drops around the critical point
even without the truncation.
The difference between lowest values of the transition frequencies
with and without the truncation
is 6.0\;\% for $N=1$ and 4.2\;\% for $N=2$,
and the difference between the optimal $\LRz$
(giving the lowest transition frequencies) with and without the truncation
is 4.5\;\% for both $N=1$ and 2.
In this way, the influence of the truncation
on the lowest transition frequencies is suppressed
at least by increasing $N$ from 1 to 2.

% \bibliographystyle{bamba_prl_em}
% \bibliography{../../../../OneDrive/bib/list,../../../../OneDrive/bib/bamba}

\end{document}